\documentclass{ws-ijmpcs}
\newcommand{\thcm}{\theta_\text{cm}}
\begin{document}

\markboth{Authors' Names}
{Instructions for Typing Manuscripts (Paper's Title)}

\catchline{}{}{}{}{}
%
%%%%%%%%%%%%%%%%%%%%%%%%%%%%%%%%%%%%%%%%%%%%%%%%%%%%%%%%%%%%%%%%%%%%

\title{Incompleteness of complete kaon photoproduction}

\author{JAN RYCKEBUSCH, TOM VRANCX}

\address{Department of Physics and Astronomy, Ghent University,
  Proeftuinstraat 86,\\ B-9000 Gent, Belgium}

%\address{Group, Laboratory, Address\\
%City, State ZIP/Zone, Country\\
%second\_author@domain\_name}

\maketitle

\begin{history}
\received{Day Month Year}
\revised{Day Month Year}
\end{history}

\begin{abstract}
A possible roadmap for reaching a status of complete information in
$\gamma p \rightarrow K^ {+} \Lambda$ is outlined. 
\end{abstract}

\ccode{PACS numbers: 11.80.Cr, 13.60.Le, 24.10.−i, 25.20.Lj}

\section{Introduction}	
Thanks to recent technological advances in producing high-quality
polarized monochromatic photon beams, and in developing polarized
nucleon targets, it becomes possible to measure a sufficiently large
amount of single- and double-polarization observables in pion and kaon
photoproduction from the nucleon. As a result, a status of complete
quantum mechanical information of meson photoproduction comes within
reach.  Measurements are complete whenever they enable one to
determine unambiguously all amplitudes of the underlying reaction
process at some specific kinematics.

\begin{figure}[pb]
\centerline{\psfig{file=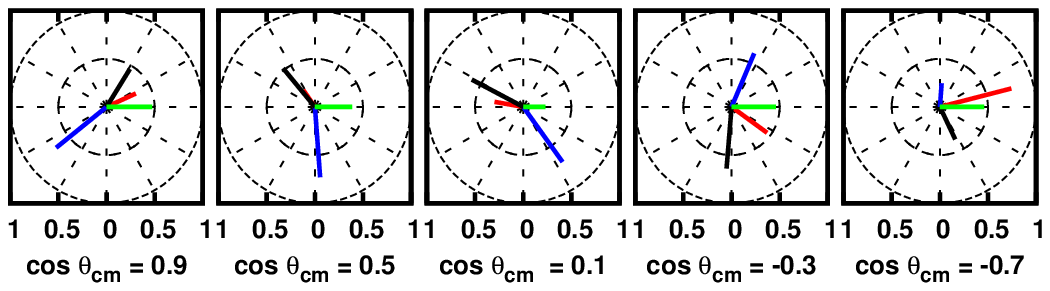,bb=51 122 354 210,clip,width=\textwidth}}
\vspace*{8pt}
\caption{The RPR-2011 predictions for the NTA (the definitions are
  given in Ref.~\protect \cite{TomComp}) for the $\gamma p \rightarrow K^{+}
  \Lambda$ reaction at $W=1900$~MeV. The NTA are displayed as vectors:
  $a_{1}=r_{1} e ^ {i \alpha _1} $ (red), $a_{2}=r_{2} e ^ {i \alpha
    _2}$ (blue), $a_{3}=r_{3} e ^ {i \alpha _3}$ (black), $a_{4}=r_{4}
  $ (green).  We adopt the convention that $\alpha_4$=0.}
\label{f1}
\end{figure}

We consider the $\gamma p \rightarrow K^ {+} \Lambda$ reaction as a
prototypical example of pseudoscalar-meson photoproduction from the
proton. The transversity amplitudes $b_{i} (i=1,2,3,4)$ express the
transition matrix elements in terms of the $p$ and $\Lambda$ spinors (with
quantization axis perpendicular to the reaction plane) and of linear
photon polarizations $(J_x,J_y)$.  We propose \cite{TomComp} to use
normalized transversity amplitudes (NTA) $a_{i}\equiv
\frac{b_{i}}{\sqrt{\sum_{i} \left| b_{i} \right| ^ {2}}}$ to perform
an amplitude analysis of the single- and double- polarization
observables. The NTA provide complete information after determining
the differential cross section. The corresponding polarization
observables can be expressed in terms of linear and nonlinear
equations of bilinear products of the $a_{i}$. For a
given kinematical setting determined by the meson angle $\thcm $ and the
invariant mass $W$, the $a_{i} = r_i e ^{i \alpha _i }$ are fully
determined by six real numbers conveniently expressed as three real moduli
$r_{i} $ and three real relative phases $\alpha_i - \alpha_4$.  All
observables are invariant under a transformation of the type $\alpha
_i \rightarrow \alpha _i + \beta$, with $\beta$ an arbitrarily chosen
overall phase.  In Fig.~\ref{f1} we show predictions for the NTA at
$W=1900$~MeV and various $\thcm$. The adopted model for $\gamma p
\rightarrow K^ {+} \Lambda$ is the Regge-plus-Resonance (RPR) approach in
its most recent version RPR-2011 \cite{LesleyPRC,LesleyPRL}. The model
has a Reggeized $t$-channel background and the $s$-channel resonances
$S_{11}(1650)$, $F_{15}(1680)$, $P_{13}(1720)$, $D_{13}(1900)$,
$P_{13}(1900)$, $P_{11}(1900)$, and $F_{15}(2000)$. The RPR approach provides a
low-parameter framework with predictive power for $K^ {+}$ and $K^{0}$
photoproduction on the proton and the neutron \cite{PieterNPA}.

%Over the years possible strategies to turn complete
%measurements into efficient tools to learn about the underlying
%dynamics of meson photoproduction have been proposed. Complications
%arise due to the non-negligible uncertainties in real data. In
%addition, there are issues related to the fact that some observables
%are linked to bilinear products of amplitudes by means of nonlinear
%equations.

\section{Extracting the moduli and phases from $\gamma p \rightarrow K^{+}
  \Lambda$ data}
\begin{figure}[pb]
\centerline{\psfig{file=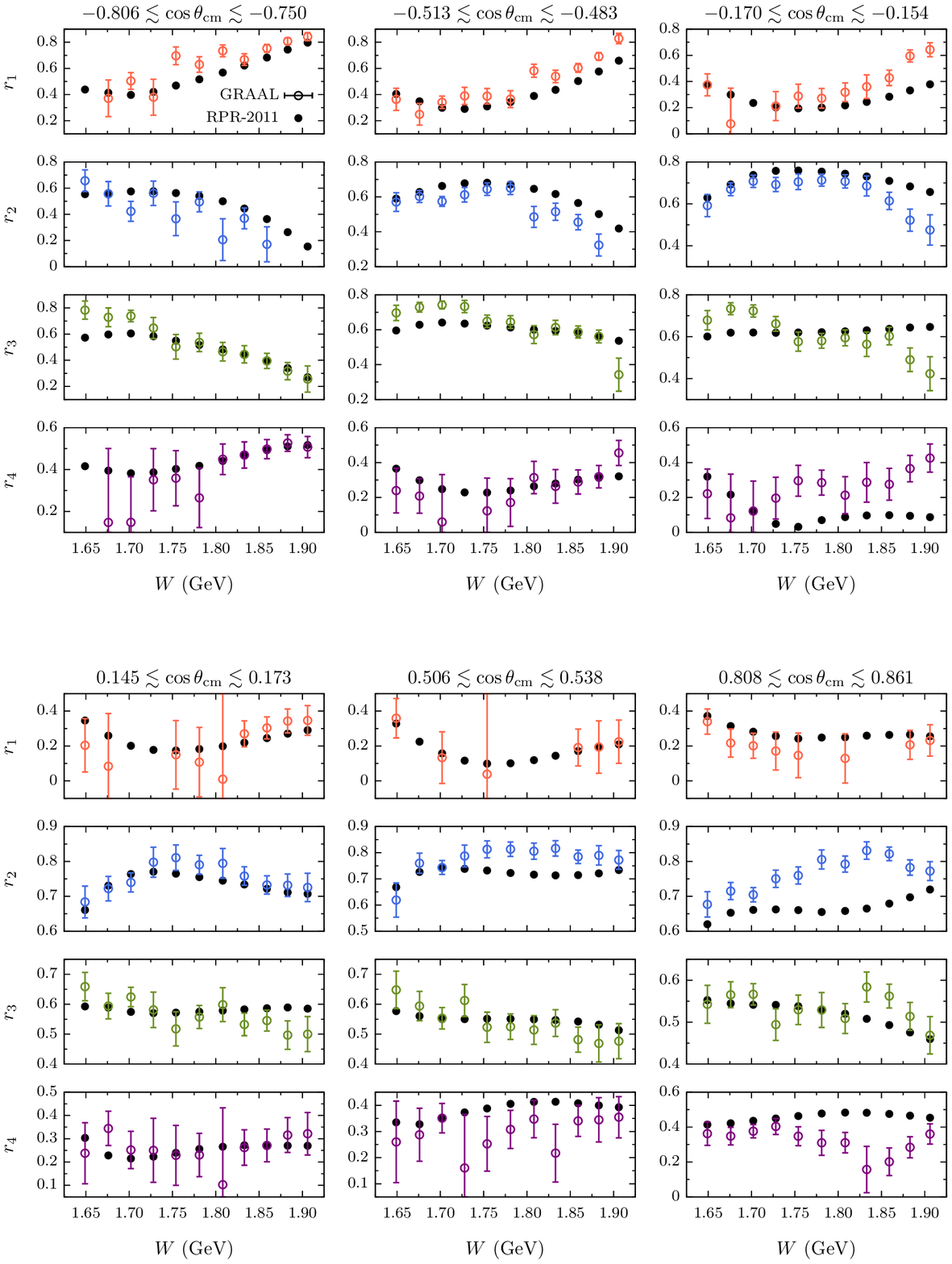,bb=51 113 545 427,clip,width=\textwidth}}
\vspace*{8pt}
\caption{The energy dependence of the moduli $r_i$ of the normalized
  transversity amplitudes for the $\gamma p \rightarrow K^+ \Lambda$
  reaction.  The data are extracted from the GRAAL results for the
  single-polarization observables reported in
  Ref.~\protect \cite{GRAALresults}. The dots are the bin-centered RPR-2011
  predictions. }
\label{f2}
\end{figure}

An obvious advantage of using the transversity amplitudes is that linear
equations connect the moduli $r_i$ of the NTA to the
single-polarization observables $\left\{ \Sigma, T, P \right\}$
\begin{equation}
\left\{
\begin{array}{lll}
r_1 &=& \frac{1}{2}\sqrt{1 + \Sigma + T + P}, \\ 
r_2 &=& \frac{1}{2}\sqrt{1 + \Sigma - T - P}, \\
r_3 &=& \frac{1}{2}\sqrt{1 - \Sigma - T + P}, \\ 
r_4 &=& \frac{1}{2}\sqrt{1 - \Sigma + T - P} \; .
\end{array}
\right.
\label{eq:1}
\end{equation}
Accordingly, a measurement of $ \left( \Sigma, T, P \right)$ at given
$(W, \cos\thcm)$ allows one to infer the moduli $r_i (W, \cos\thcm)$
of the NTA.  The GRAAL collaboration provides $p(\gamma, K^
{+})\Lambda$ data for $ \left\{ \Sigma, T, P \right\}$ at 66 $(W,
\cos\thcm)$ combinations in the ranges $1.65 \lesssim W \lesssim
1.91$~GeV ($\Delta W \approx 50$ MeV) and $-0.81 \lesssim \cos\thcm
\lesssim 0.86$ ($\Delta \cos\thcm \approx 0.3$).  Figure~\ref{f2}
shows the extracted $r_i$ at three $\thcm$ intervals along with
the RPR-2011 predictions. For a few kinematic points the
$r_{i}$ could not be retrieved from the data. This occurs whenever one
or more arguments of the square roots in Eq.~(\ref{eq:1}) become
negative due to finite experimental error bars. The RPR-2011 model
offers a fair description of the $W$ dependence of the extracted
$r_{i}$ except for the most forward angles at $W \approx
1.85$~GeV. Furthermore, the data confirm the predicted dominance of
the $r_{2}$.  

Inferring the NTA phases $\alpha _i$ from data requires measured
double asymmetries. Complete sets of the first kind, which involve
seven observables (e.g. $\{{\Sigma, P, T}, C_x,O_x,E,F\}$), lead to the
following set of nonlinear equations for the phases
\cite{chiang,TomComp}
\begin{equation}
\left\{
\begin{array}{lll}
{r_1r_4}\sin\delta_1 + {r_2r_3}\sin\Delta_{23} &=& -\frac{C_x}{2}, \\
{r_1r_4}\cos\delta_1 + {r_2r_3}\cos\Delta_{23} &=& +\frac{O_x}{2},  \\
{r_1r_3}\cos\Delta_{13} - {r_2r_4}\cos\delta_2 &=& +\frac{E}{2}, \\
{r_1r_3}\sin\Delta_{13} + {r_2r_4}\sin\delta_2 &=& -\frac{F}{2},  \\
\delta_1 + \Delta_{23} - \delta_2 - \Delta_{13} &=& \phantom{+}0 
\end{array}
\right.
\end{equation}
where $\delta _i \equiv \alpha _i - \alpha _4 $ and $\Delta _{ij} =
\delta _{i} - \delta_{j}$. Solutions to the above set of nonlinear
equations gives the phases $\left(\delta_1, \delta_2,\Delta_{13},
  \Delta_{23} \right)$ for given moduli $ \left(r_1,r_2,r_3,r_4
\right)$. We stress that single-polarization observables are part of
any complete set as they provide the information about the
moduli. Double polarization observables are required to get access to
the phases. In all practical situations one has $ \delta_1 +
\Delta_{23} - \delta_2 - \Delta_{13} \approx 0 $. Finite error bars
introduce a bias for the choices made with regard to the reference
phase (here, $\alpha _{4}$) for the above equations.  A consistent set
of estimators $\widetilde{\delta}_i^{\;\alpha_j}$ for the independent
phases (insensitive to choices made with regard to the reference
phase) has been proposed in Ref.~\cite{TomComp}.

To date, the published double polarization observables for $\gamma p
\rightarrow K^ {+} \Lambda$ do not allow one to extract the phases of
the NTA.  We have conducted studies with pseudo-data generated by the
RPR-2011 model for $\gamma p \rightarrow K^ {+} \Lambda$.  We have
considered ensembles of 200 pseudo-data sets each containing samples
of 50 events for the asymmetries $\{\Sigma, P, T, C_x,O_x,E,F\}$. The
pseudo-data are drawn from Gaussians with the RPR-2011 prediction as mean
and a given $\sigma_{\textrm{exp}}$ as standard deviation.  The retrieved
$(r_i, \delta _i)$ do not necessarily comply with the input
amplitudes.  There are various sources of error: (i) imaginary
solutions for the moduli ; (ii) imaginary solutions for the phases; (iii)
incorrect solutions which stem from the fact that $\delta_1 +
\Delta_{23} - \delta_2 - \Delta_{13} = 0$ cannot be exactly obeyed
for data with finite errors. We find that the amount of incorrect and
imaginary solutions is much larger for the phases than for the
moduli. The frequency of finding imaginary solutions can be
dramatically reduced by improving on the experimental resolution
$\sigma_{\textrm{exp}}$.

%Insolvability $\eta = \eta_\text{imaginary} + \eta_\text{incorrect}$ 
% is a measure for the fraction of complete measurements which does not result in a successful determination of NTA. 

\section{Conclusions}

We have sketched a possible roadmap for reaching a status of complete
information in pseudoscalar-meson photoproduction. We suggest that the
use of transversity amplitudes is tailored to the situation that
experimental information about $\left( \Sigma, P, T \right)$ is more
abundant (and most often more precise) than for the double
polarization observables. Linear equations connect $\left\{ \Sigma, P,
T \right\}$ to the moduli $r_i$ of the NTA. An analysis of $\{\Sigma,
T, P\}$ data for $\gamma p \to K^+\Lambda$ from GRAAL allowed us to
extract the $r_i$ in the majority of considered $\left(W,\cos \thcm
\right)$ combinations.  Extracting the NTA independent phases is far
more challenging as they are connected to the double asymmetries by
means of nonlinear equations. It has been suggested \cite{Lothar}
that over-complete sets which involve more than seven polarization
observables may provide a solution to tackle the problem of extracting
the relative phases of the amplitudes from the data.

\section*{Acknowledgments}

This work is supported by the Research Council of Ghent
University and the Flemish Research Foundation (FWO
Vlaanderen).

%References are to be listed in the order cited in the text in Arabic
%numerals.  They should be listed according to the style shown in the 
%References. Typeset references in 9 pt roman.

%\begin{thebibliography}{000} %for 3 digits
%\begin{thebibliography}{00}  %for 2 digits


\begin{thebibliography}{0}    %for 1 digit
%
\bibitem{TomComp}
T. Vrancx, J. Ryckebusch, T. Van Cuyck, P. Vancraeyveld, 
{\it Phys. Rev. C} {\bf 87}, 055205 (2013). 
%
\bibitem{LesleyPRC} L. De Cruz, J. Ryckebusch, T. Vrancx, P. Vancraeyveld,
{\it Phys. Rev. C} {\bf 86}, 015212 (2012).
%
\bibitem{LesleyPRL}
L. De Cruz, T. Vrancx, P. Vancraeyveld, J. Ryckebusch,
{\it Phys. Rev. Lett.} {\bf 108}, 182002 (2012).
%
\bibitem{PieterNPA}
P. Vancraeyveld, L. De Cruz, J. Ryckebusch, T. Vrancx, 
{\it Nucl. Phys.}  {\bf A897}, 42 (2013).
%
%
\bibitem{GRAALresults}
A. Lleres {\it et al.} (GRAAL Collaboration), {\it Eur. Phys. J. } {\bf A31}, 79
(2007) and {\bf A 39}, 149 (2009).
%
\bibitem{chiang}
Wen-Tai Chiang and F. Tabakin, {\it Phys. Rev. C} {\bf 55}, 2054 (1997).

\bibitem{Lothar}
R. L. Workman, M. W. Paris, W. J. Briscoe, L. Tiator,
S. Schumann, M. Ostrick, and S. S. Kamalov, {\it Eur. Phys. J.}
{\bf A47}, 143 (2011).


\end{thebibliography}
\end{document}